 \documentstyle[preprint,revtex,epsf]{aps}
\tightenlines
\begin{document}
\begin{title}
The Reality and Measurement of the Wavefunction
 \end{title}

\author{W. G. Unruh}
\begin{instit}
 CIAR Cosmology Program\\
Dept. of Physics\\
University of B. C.\\
Vancouver, Canada V6T 2A6\\
\end{instit}

{}~

{}~

\begin{abstract}
Using a simple version of the model for the quantum measurement of a
two level system, the contention of Aharonov, Anandan, and Vaidman
 that one must
in certain circumstances give the wavefunction an ontological as well
as an epistemological significance is examined. I decide that their
argument that the wave function of a system can be measured on a
single system fails to establish the key point and that what they
demonstrate is the ontological significance of certain operators in the
theory, with the wave function playing its usual epistemological role.

\end{abstract}

Aharonov, Anandan, and Vaidman\cite{1} have recently argued that
 in addition to its usual
epistemological role, the wave function in quantum mechanics in certain
situations also has an ontological role. In other words, in addition to
acting as a device in the theory to encode the conditions (our
knowledge of the world) it must also, in certain circumstances, be
regarded as real, in the sense that one can completely determine an
unknown wavefunction of  a single system. Certainly if their claim were
true, that one could take a single system with an unknown wavefunction,
and completely determine that wave function on that single system, one
would have to accord the wave function a reality on the grounds that
that which is measurable is real. In the course of this paper I will
argue that they have failed to establish the measurability of the wave
function, and thus have failed in their attempt to demonstrate the
reality of the wave function. The argument is however subtle. Thus the
plan of this paper will be to first discuss the problem of reality in quantum
mechanics, to set stage for the question that they are trying to answer.

I will then go through their argument in some detail for a simple
system which will I hope clarify their argument. In particular I hope
it will clarify the key term in their paper, namely ``protection". I
will finally argue that they have failed to establish a key requirement
of their argument. To use their term, I will argue that ``protection"
is not an active attribute, in that one cannot protect a wave function.
Rather protection is an attribute that a system can have, and that the
wave function will play its usual epistemological role in stating the
condition that the system has that property. It is in their failure to
establish the active, rather than the passive sense of protection that
I feel their argument fails.

Quantum mechanics is a strange theory which even today, seventy years
after its invention/discovery, causes immense difficulties to students
and physicists alike. In particular, many find it difficult  to
establish a reliable quantum picture of reality. Physicists are wedded
to the
proposition that their field of study is one concerned with reality,
that their mathematical formalism is an accurate model of something
which exists in the outside world, and is not simply a free invention
of the human mind.

In classical physics the relation between the mathematical theory and
reality is relatively direct. The terms in the theory correspond in a
relatively direct way with physical reality. The dynamical variables in
the theory can be assigned values, and so can the attributes of objects
in the
world. The assigning of values to attributes of the real world in the
process of measurement can be modeled directly in the assigning of
values to the variables in the theory. The variables in the theory can
therefore be taken as having a direct correspondence with the attributes
in the world.

 Quantum mechanics on the other hand does not behave in such a neat
 manner. There seems to be no direct correspondence between the structures
in
 the theory and  the attributes of the real world. It
 is this more than anything else which has lead to the widespread
 unease with quantum mechanics as an ultimate theory of nature. There are
two separate structures in quantum mechanics, the operators on some Hilbert
space, and the vectors which live in that Hilbert space. Both of these
mathematical structures play a role in the theory, but how do they relate
to the structure of the physical world?

The operators represent the dynamical variables, and thus correspond in
the most direct way to the attributes of the physical world. However, one
cannot simply ascribe a number to an operator. The correspondence between
the values that we know physical attributes can have, and the  assigning
of values to the operators is far less straightforward than in classical
physics. The values of the operators are its eigenvalues, and one can assign
a value to an operator only when it is operating on special vectors in
Hilbert space, namely the eigenvectors. Furthermore the various operators
cannot all simultaneously have values. This feature of quantum theory is
of course well known. The state vector, or wave function thus tells us
which of the various operators and thus which of the physical attributes
can be assigned a definite value. It furthermore plays the mysterious role
of assigning probabilities to the various values that other dynamical
 variables can have.  It thus plays an epistemological role in
representing our knowledge about the system under consideration.

 But what role does it play in physical
reality? Is there some aspect of the physical world which corresponds to
the wave function, some extra aspect of the world separate from the dynamic
variables, or is it an ideal element of the
theory, with no physical reality in its own right, and existing solely in
the theory to represent our knowledge of the actual state of the physical
world? Is the the wave function
real  or is it simply a device within the theory to incorporate our knowledge
of the world, without it in itself corresponding to anything in the real
world?
In the former case it would have an ontological role, while in the latter
a purely epistemological role.

That physicists have long wanted to regard the wave function as having
an ontological role is clear. That desire underlies the unease surrounding
the ``projection postulate", and the oft heard lament that the wave function
cannot simultaneously obey deterministic dynamical equations of evolution,
and indeterministic collapse in the ill defined ``measurement" process.
As an object with an ontological role, as an object corresponding to some
element of physical reality such schizophrenic behaviour is highly
unsatisfactory.
On the other hand if the wave function is simply a tool by which we encode
our knowledge into the theory, the change of the wavefunction under a change
in knowledge is perfectly rational. It leaves one, however with uncomfortable
questions about how knowledge differs from other physical processes.

 AAV try to answer the question about the reality of the wave function by
asking whether or not the wave function
is measurable. If not just the dynamic variables, but the wave function
itself can be measured, then the wave function must correspond to an aspect
of reality, must have an ontological as well as an epistomelogical aspect
within the theory.
Thus the question they ask is whether one can start with a single system
and on that system determine what its
wave function is?  For example, if one can determine the expectation value
of a sufficiently
large number of operators,  one can determine the wave function. Now, the
expectation value can be determined if we start with a sufficiently large
ensemble of systems, all in the same state. But this would at best give
the wave function a role in ensembles, not in single systems.

 To give it
a true ontological meaning, one would like the wave function to have a
physical significance, to be measurable, on a single system. Thus the question
is ``Can one take a system  in some initial {\it unknown} state and completely
determine that state by a sequence of measurements on that single system?"   If
one  can,
then one would
be forced to the position that the the wave function of a single system is
measurable
 and that the wave function therefor was a real attribute of that single
system. This
is what AAV claim. What they show however, is that, if the wavefunction is
known
beforehand to be the eigenstate of the unknown Hamiltonian of that system then
it
is possible to determine the properties of that eigenstate.  What their
analysis
shows is that one can determine some of the properties of an unknown
Hamiltonian
on a single system, if one knows that the system is in an eigenstate of that
operator.

Let us therefor analyze the AAV contention that such a measurement is possible
under certain conditions.
To do so, I will concentrate on a specific, simple, system, namely a two
level (spin ${1\over 2}$) system. Can one determine what the wave function
of such a system is on a single exemplar of that system? AAV claim the answer
to be yes, if one can , to use their word, protect the wave function.

The difficulty with the usual measurement procedure is that the wave function
in general changes as the result of measurements. In the view of measurement
as a primitive of the theory, this change is just the projection postulate,
the reduction of the wave function. In the view of measurement as a physical
process, it is the result of the interaction of the measuring apparatus
with the system being measured.
Thus after a single measurement, the system is no longer in the state it
began with, and there is no way in which one can recover that original
state.
Furthermore a single measurement of any time is not sufficient to determine
the state of the system.

However, they argue that if one can "protect" the wave function, ie interact
with the system in such a way that its wave function remains unchanged
after the measurement, but still affects the measuring apparatus, that
then one can make a succession of such measurements on the system, and
finally completely determine its wave function.

The key to their ``protection" was to realize that if a system is in an
energy eigenstate, and if the interaction between the system and the rest
of the world was adiabatic, that then the wave function after the measurement
would still be that same energy eigenstate.  The wave function would not
have changed even though the system would have interacted with the rest
of the world, and even though its interaction could have changed that outside
world in a manner that depended on the specific state that the system was
in. Thus one could measure the system ( ie determine its effect on the
outside world, and thus the state it was in) without changing the wave
function of the system. This allows one to make repeated measurements on
the system, and completely determine the state of the system.

To make these ideas clear, let us use a simple two level system as our
system of interest. It is assumed to be in some pure state $|\psi_0\rangle$.
Now,
for a two level system any pure state can be represented by a three dimensional
vector $\vec\rho$, such that $P={1\over 2} (1+\vec\rho \cdot \vec\sigma)$
is the projection operator onto that pure state. Thus given an unknown
wave function $|\psi_0\rangle$, if we can determine its associated vector
$\vec\rho$,
we will have completely determined that wave function.

Let us now assume that we can arrange that this system has $H_0=P$ as its bare
Hamiltonian. This is the crucial assumption. It is exactly this condition
 which is the condition of ``protection" introduced by AAV.

 Our measuring apparatus will be a free, infinitely massive
particle. The ``infinitely massive" assumption is made so that the the
 $p^2/2m$ free Hamiltonian for the measuring apparatus can be neglected.
 The coupling to  to the measuring apparatus will be via an interaction of the
form
$$
H_{int}= \epsilon (t) \vec x\cdot {\bf L}\cdot \vec\sigma
$$
where ${\bf L}$ is a known matrix, and $x$ is the position operator of the
particle.
$\epsilon(t) $ is a time dependent coupling which is assumed to be slowly
varying. The full Hamiltonian for the system is then given by
$$
H= {1\over 2} (1+\vec \rho\cdot\vec\sigma)
+\epsilon(t) \vec x \cdot {\bf L} \cdot \vec\sigma
$$

Now, the Schroedinger equation is simply solved for this Hamiltonian.
The position $\vec x$ is a constant of the motion, so that we can
expand the states of the apparatus in eigenstates of the position,
 $|x\rangle$. The Schroedinger equation for the system plus apparatus,
 with the apparatus in an eigenstate of position, is then
\begin{equation}
-i{\partial\over\partial t}|\psi_x(t) \rangle= {1\over 2} \left(1+
 (\vec \rho+2\epsilon \vec x \cdot {\bf L})\cdot \vec\sigma\right) |\psi_x(t)>
\end{equation}
where the full wave function for the system plus apparatus is given by
$$|\Psi(t)\rangle = \int |\psi_x(t)\rangle|x\rangle dx$$.

The largest instantaneous eigenvalue of this operator on the right
of this reduced Schroedinger equation is given by
\begin{equation}
E(t)= {1\over 2} \left(1+
\sqrt{(\vec \rho+ 2\epsilon \vec x\cdot{\bf L})
\cdot (\vec \rho+ 2\epsilon \vec x\cdot{\bf L})}\right)
\end{equation}
and the projection operator onto this eigenvector $|E(t)\rangle$ is given by
\begin{equation}
P_E(t)
= {1 \over 2}
\left(1+ {(\vec\rho+2\epsilon(t) \vec x\cdot{\bf L})\over E(t)}\cdot
\vec\sigma\right)
\end{equation}

By the adiabatic theorem, the state of the system will evolve as the
eigenvector of this projection operator, which I will label by $|E(t)>$.
 This will be true as long as the {\bf direction} of the unit vector
\begin{equation}
\hat n(t)={(\vec\rho+2\epsilon(t) \vec x\cdot{\bf L})\over E(t)}
\end{equation}
does not change rapidly in time. ($ |\dot {\hat n}|\ll E(t)$).
 If this approximation is valid, then
\begin{equation}
|\Psi\rangle = \int e^{-i\int^t E(t') dt'}
 |E(t)\rangle |x\rangle \langle x|\phi_0\rangle dx
\end{equation}
where $|\phi_0\rangle$ is the initial state of the measuring apparatus.
Now, let us assume that the the values of $\vec x$ for which the amplitude for
the measuring apparatus
is non zero are small (i.e., $\phi(x)=0$ for large $x$). We can then write
\begin{equation}
E(t)\approx 1 + \epsilon(t) \vec x \cdot{\bf L} \cdot \vec \rho~ +O((\epsilon
x)^2)
\end{equation}
Retaining the first term, and looking at the system after $\epsilon(t)$
has dropped back to zero again, we find that
\begin{equation}
|\Psi\rangle
\approx |\psi_0\rangle \int e^{-i\int_{-\infty}^{\infty} \epsilon(t) dt
{}~(\vec x\cdot {\bf L}\cdot\vec \rho) } ~
\langle x|\phi_o\rangle |x\rangle d^3x
\end{equation}
But the exponential is just the momentum space displacement operator. Thus
if the original measuring apparatus had a wave function of the form
$\phi(\vec p)= \langle \vec p|\phi\rangle$,
the state of the measuring apparatus after the measurement is
$\phi(\vec p+\int \epsilon {\bf L}\cdot\vec \rho)$.
The momentum of the  measuring apparatus  has been displaced by
an amount just proportional
to ${\bf L}\cdot\vec \rho$, and the wave function of the system
plus apparatus after the measurement is a
product state with the system still in its original state.

Now, we can write ${\bf L}\cdot\vec \rho$ as
\begin{equation}
{\bf L}\cdot\vec \rho = \langle \psi|{\bf L}\cdot \vec \sigma |\psi\rangle,
\end{equation}
 since in the assumed initial state, $\vec\rho=\langle
\psi|\vec\sigma|\psi\rangle$.
and we have that the measuring apparatus has responded to the expectation
value of an operator in the state of the system. By making a sufficient
number  measurements on the system, with various values of ${\bf L}$,
one can therefor completely determine $\vec \rho$ and therefor the original
wave function for the system. Because the final state of the system
 is identical with the initial state, one is furthermore able to
 make as many measurements as one desires, with as many different
measuring apparatuses and as many different {\bf L} as desired.
An interesting but not significant fact about the specific example
 analyzed here is that  if we take {\bf L} to be the identity
matrix (or any other non-singular matrix), a single measurement of the
change of the momentum of the
apparatus will give us, to within some error determined by the spread
of the initial momentum of
the apparatus, the value of $\vec \rho$.  Thus ,
their argument in a nutshell is, that by protecting the wave function,
 ie, choosing the free Hamiltonian of the system so that the initial
 state of the system is  an eigenfunction of the Hamiltonian, and
thus protecting it,  the wave function can be determined and
measured on a single system ( and in this case, with a single measurement).

The crucial step is the protection postulate. Can a wave function be protected
or is it rather a feature that the wave function has, like being the eigenstate
of some operator. Is one measuring the wave function in the above procedure,
or is one measuring one of the dynamic variables of the system.
for example, it is well known that one can measure various properties of
a single system. In particular, one can measure the properties of the
Hamiltonian
of a single system.  The detailed Hamiltonian of a single atom enclosed
within a Penning trap for example can be measured with arbitrary accuracy
on that single system. The crucial point about the measurement of the wave
function is whether or not one can measure an arbitrary wave function
of a given system, and this they fail to show. In the above example, the
arbitrariness of the wavefunction was encoded in the arbitrariness of the
vector $\vec\rho$. The assumption was made that the Hamiltonian of the
system had the same vector $\rho$ as did the wave function. If one could
take an arbitrary wave function and without knowing what that wave function
was, one could choose  the Hamiltonian for the system so that it was given
by that same $\rho$, then one could genuinely say that one had measured
the wavefunction by the above technique. However, the situation we have is that
the Hamiltonian for the system is given by some unknown $\rho$, and we can
choose
the state so that it is described by the same $\rho$, ie is an eigenfunction of
 that same Hamiltonian.

One may be able to ensure that the particle is in the eigenstate of the unknown
Hamiltonian (although one's ability to do even this may be questionable,
especially if the state in question is not the ground state of that unknown
Hamiltonian), but one cannot ensure that the Hamiltonian has the given but
unknown state as its eigenstate. To do the latter, to take the system in an
unknown state, and to then choose a Hamiltonian for the system so that that
unknown state is its eigenstate  would mean that the equation
of evolution of the wave function would be non-linear.
 The Hamiltonian driving the evolution would be a function of the
initial state of the system. If we could carry out that procedure for
our example,
in the Schroedinger representation, one would have
\begin{equation}
i{\partial \psi\over \partial t} = {1\over 2} (1+\vec \rho(\psi)\cdot\vec\sigma
)\psi
\end{equation}
Because the Hamiltonian depends on the wave function itself, the
evolution is no longer a linear evolution, but is highly
non-linear.
Instead one has
\begin{equation}
i{\partial \psi\over \partial t} = {1\over 2} (1+\vec \rho_0\cdot\vec\sigma
)\psi
\end{equation}
where $\rho_0$ is some constant independent of $\psi$.

This point lies at the heart of their paper. If one comes upon the system after
 the initial selections have been made, one could say that all one has is a
state
 and a Hamiltonian. Both are unknown, but it is known that the state is an
eigenstate of the Hamiltonian. What does it matter how they got that way? What
 does it matter whether the physicists began with an unknown Hamiltonian and
forced the system to be in an eigenstate of the hamiltonian, or one started
with an unknown state, and  chose the Hamiltonian so be the projection operator
onto that state say? The end result is identical. However, if one is to regard
the wave function as real,  as having an ontological significance for that
single system, the two are crucially different. We can force the wave function
to be an eigenstate of an operator, we cannot force the operator to have the
unknown wavefunction as its eigenstate. It is precisely the impossibility of
the latter that leads to the conclusion that the wave function is not real,
does not have the necessary permanence and independence to be considered as
real.

What happens if we present their system with the spin half system in a truly
 unknown state? The system has some (unknown) Hamiltonian, which we cannot
force to have the unknown state as its eigenstate. Thus the bare Hamiltonian
is given by
\begin{equation}
H_0={1\over 2} (1+\rho_0\cdot \sigma)
\end{equation}
where $\vec\rho_0$ is unknown but is {\bf not} equal to
 $\rho=<\psi|\vec\sigma|\psi>$.
Given this as the Hamiltonian for the system, and given the same coupling
to the apparatus, one finds that the initial wave function  for the system will
be
\begin{equation}
|\psi_0\rangle= \cos(\theta) |+\rangle +\sin(\theta) |-\rangle
\end{equation}
where
$$
\cos(\theta/2)= \vec\rho\cdot\vec\rho_0
$$
and the phase of $|-\rangle$ has chosen appropriately ( it will not
 enter into the following analysis).
$|+\rangle ,|-\rangle $ are the eigenstates of
$\vec\rho_0\cdot\vec\sigma$ After the interaction, the
final wave function is given by
\begin{equation}
|\Psi\rangle
= \cos(\theta)|+\rangle \phi(\vec p-\int \epsilon dt {\bf L}\cdot\vec\rho)
+ \sin(\theta)|-\rangle \phi(\vec p)
\end{equation}
If $\int\epsilon dt{\bf L}\cdot \vec\rho_0$ is sufficiently large and the
wave function $\phi(\vec p)$ sufficiently peaked,
then a measurement of $\vec p$
after the interaction will tell one that the system was in the $|+\rangle $
state with probability $\cos^2(\theta)$
and in the $|-\rangle$ state with probability $\sin^2(\theta)$.
 Ie, the measurement
 envisaged by AAV leads to the standard result that the
measurement  gives us one of the two possible outcomes for the polarisation
 of the spin half system along the direction of $\rho_0$ with the
usual probabilities.
Their protected measurements are simply another form of the standard quantum
measurement theory. They measure the polarisation of the particle
along the direction of the (possibly unknown) direction $\rho_0$.

Note that this measurement procedure is somewhat different from the standard
 von Neumann analysis of the measurement process. There the measurement is of
 the operator of the system which couples the measuring apparatus to system.
In this problem that would be the vector operator  ${\bf L}\cdot\vec\sigma$.
 Rather the measurement is one of the free Hamiltonian of the system, $H_0$.
But
such situations were already well known from the standard analysis of the Stern
Gerlach
experiment in which the particle is measured to be in the eigenstates of
spin along the constant magnetic field ( which does not couple to the
translational
degrees of freedom of the particle) rather than the spin along the direction
of the inhomogeneous field, which does couple to the translational degrees
of freedom by applying a force to the particle. In fact as the above analysis
makes clear, our system and measuring apparatus of the standard Stern Gerlach
 experiment\cite{2}, with ${\bf L}$ just the matrix $\vec\nabla \vec B$, and
 the measuring apparatus being the translational degrees of freedom of the
particle.

 I commented that the deflection of the measuring
apparatus was proportional to ${\bf L}\cdot\vec\rho$, which could be written
as the expectation value of the operator ${\bf L}\cdot \vec \sigma$ in the
special
``protected" state. This is one of the key points which led
AAV to conclude that
 this process can be considered to be a measurement of the state of the system.
However, the change in momentum of the apparatus
 can also be written as being proportional to
 $Tr H_0 {\bf L}\cdot \vec\sigma$. In fact when we choose the state to be
different from the eigenstate of the Hamiltonian, we realize that this is
a better expression for the deflection, because it applies for an arbitrary
initial state.
The deflection remains the same regardless of the state
 of the system, as long as $H_0$ remains the same. Thus, one would be more
accurate to regard the  the deflection is a property
 not of the state of the system, but of
the dynamic operators which define the system and the measuring apparatus.
One is not measuring a property of the state of the system, namely the
expectation value of some operators. Rather one
is measuring properties of the dynamical operators. It is they, not the
wave function which have ontological property. The wave function, in their
analysis, plays its usual epistemological role in telling us that we are
in an eigenstate, that the system has a certain definite energy. Just because
we do not happen to know exactly what the Hamiltonian is that the system is in
the eigenstate of, does not give the wave function any ontological property
that it would not have if we knew what the Hamiltonian was.

Thus, I have shown  that the analysis of AAV is crucially flawed in
demonstrating
 the reality of the wavefunction. If a system has an unknown Hamiltonian, and
if
one knows that the state of the system is in an
eigenstate of that unknown Hamiltonian (ie, the wave function plays
its usual epistemological role in designating knowledge we have
about the system), then one can determine that eigenstate--- ie
 that property of the unknown Hamiltonian-- on that single system.
  However, if the system is in a completely unknown state, then they
 have not demonstrated that they can protect that state, and there is
 no measurement or sequence of measurements which will tell us what
that state is. I.e., an unknown state is not measurable, and thus is
not real or ontological.

\acknowledgments
I would like to thank J. Anandan for sending me a copy of the paper before
publication, and Leslie Ballentine for his helpful comments on previous
 versions of this manuscript. I would also like to thanks the Canadian
Institute for Advanced
Research and the Natural Science and Engineering Research Council for support
while this work was being done.

\end{document}